\documentclass[conference]{IEEEtran}
\IEEEoverridecommandlockouts
\usepackage{cite}
\usepackage{amsmath,amssymb,amsfonts}
\usepackage{algorithmic}
\usepackage{graphicx}
\usepackage{textcomp}
\usepackage{xcolor}
\def\BibTeX{{\rm B\kern-.05em{\sc i\kern-.025em b}\kern-.08em
    T\kern-.1667em\lower.7ex\hbox{E}\kern-.125emX}}

\ifCLASSOPTIONcompsoc
\usepackage[caption=false, font=normalsize, labelfont=sf, textfont=sf]{subfig}
\else
\usepackage[caption=false, font=footnotesize]{subfig}

\usepackage{xspace}
\usepackage{threeparttable}
\usepackage{cleveref}
\usepackage{booktabs} % For formal tables
\usepackage{multirow}

\usepackage{amsmath}

\usepackage{adjustbox}
\usepackage{color, colortbl}
\usepackage{changepage}
\usepackage[normalem]{ulem}

\usepackage{url}

\usepackage{color,soul} % For commenting & highlighting

\usepackage{comment}

\usepackage[shortlabels]{enumitem}

\newcommand\blfootnote[1]{%
	\begingroup
	\renewcommand\thefootnote{}\footnote{#1}%
	\addtocounter{footnote}{-1}%
	\endgroup
}

\newcommand{\dsagift}{DXOR-GIFT\xspace}
\newcommand{\slvgift}{SXOR-GIFT\xspace}
\newcommand{\dsa}{DXOR\xspace}
\newcommand{\slv}{SXOR\xspace}

\newcommand{\journal}{0} % For selecting the text for the journal

\begin{document}

\title{Memristor-Based Lightweight Encryption}

\author{\IEEEauthorblockN{Muhammad Ali Siddiqi\IEEEauthorrefmark{1}\IEEEauthorrefmark{4}, Jan Andrés Galvan Hernández\IEEEauthorrefmark{1}, Anteneh Gebregiorgis\IEEEauthorrefmark{1}, Rajendra Bishnoi\IEEEauthorrefmark{1},\\ Christos Strydis\IEEEauthorrefmark{1}\IEEEauthorrefmark{4}, Said Hamdioui\IEEEauthorrefmark{1}\IEEEauthorrefmark{3} and Mottaqiallah Taouil\IEEEauthorrefmark{1}\IEEEauthorrefmark{3}}
\IEEEauthorblockA{\IEEEauthorrefmark{1}Quantum and Computer Engineering Department, Delft University of Technology, The Netherlands\\
\IEEEauthorrefmark{4}Department of Neuroscience, Erasmus Medical Center, Rotterdam, The Netherlands\\
\IEEEauthorrefmark{3}Cognitive IC B.V., Delft, The Netherlands\\
\{m.a.siddiqi, a.b.gebregiorgis, r.k.bishnoi, c.strydis, s.hamdioui, m.taouil\}@tudelft.nl, jan-andres@live.nl
}
}

\maketitle

\blfootnote{This is the accepted version of the article that was published in the 2023 26th Euromicro Conference on Digital System Design (DSD). DOI 10.1109/DSD60849.2023.00092.}

\begin{abstract}

Next-generation personalized healthcare devices are undergoing extreme miniaturization in order to improve user acceptability. However, such developments make it difficult to incorporate cryptographic primitives using available target technologies since these algorithms are notorious for their energy consumption.
Besides, strengthening these schemes against side-channel attacks further adds to the device overheads.
Therefore, viable alternatives among emerging technologies are being sought.
In this work, we investigate the possibility of using memristors for implementing lightweight encryption.
We propose a 40-nm RRAM-based GIFT-cipher implementation using a 1T1R configuration with promising results; it exhibits roughly half the energy consumption of a CMOS-only implementation.
More importantly, its non-volatile and reconfigurable substitution boxes offer an energy-efficient protection mechanism against side-channel attacks.
The complete cipher takes 0.0034 mm$^2$ of area, and encrypting a 128-bit block consumes a mere 242 pJ.

\end{abstract}

\begin{IEEEkeywords}
Memristor, hardware security, lightweight encryption, side-channel attack, GIFT cipher, 1T1R
\end{IEEEkeywords}

\section{Introduction}
\label{sec:introduction}

In recent years, small-form-factor edge devices have been considered as crucial components for next-generation personalized healthcare through medical body area networks (MBANs)~\cite{siddiqi2022improving}. Naturally, due to the sensitive nature of these devices, security and privacy become a critical concern, which require proper incorporation of cryptographic primitives and secure communication protocols.
However, due to the ultra-resource-constrained nature of these devices, such as \textit{mm-sized} neural implants, security primitives implemented using available CMOS technology become too costly for integration. The problem is further compounded when adding additional protection mechanisms to protect the security mechanisms themselves against side-channel attacks.
Therefore, new emerging technologies need to be explored to address this issue.

New memristive technologies have recently been employed in implementing cryptographic primitives~\cite{lv2021application,wang2019finite,yang2021cryptographic,sun2022physical,cai2022hyperlock,dai2021audio,du2021low,pang2019memristors,cambou2020cryptography,james2019overview,galvan2022memristive} because of their energy efficiency and ability to protect against side-channel attacks.
%https://www.weebit-nano.com/technology/reram-advantages/
However, to the best of our knowledge, there are no works that attempt to \textit{simultaneously} take advantage of these characteristics for \textit{block encryption}, which is the preferred type of encryption for resource-constrained devices.

In this work, we attempt to explore the use of memristors in constructing a lightweight block cipher that \textit{also} offers side-channel protection at no significant additional cost. We first explore lightweight block ciphers in literature and select a suitable candidate (i.e., GIFT) to showcase the potential effectiveness of memristors in securing ultra-resource-constrained edge devices. We then implement the selected cipher using a 1T1R memristive crossbar. In essence, this work provides the following contributions:

\begin{itemize}
    \item Restructuring of a round of GIFT encryption (i.e., substitution, permutation and round-key addition) in order to allow RRAM-crossbar implementation. This makes it possible to execute a round in a \textit{single read} operation and to \textit{reuse} the same hardware for multiple (40) rounds.
    \item Exploration and evaluation of the design choices for implementing XOR functionality that is used in round-key addition.
    \item A 40-nm implementation of the GIFT cipher using a 1T1R RRAM configuration.
\end{itemize}

The rest of the paper is organized as follows:
Section~\ref{sec:background} provides a brief background and an overview of related works. Section~\ref{sec:towards-mem-based-encryption} discusses a suitable cipher that can be used to showcase the utility of memristors in lightweight encryption. Section~\ref{sec:proposed-solution} explains our proposed scheme followed by results in Section~\ref{sec:results}. We draw overall conclusions in Section~\ref{sec:conclusions}.

\section{Background}
\label{sec:background}

\subsection{Memristor and RRAM background}

\begin{figure}[!t]
    \centering
    \includegraphics[trim={0cm 0.5cm 0cm 0.2cm},clip,scale=0.47]{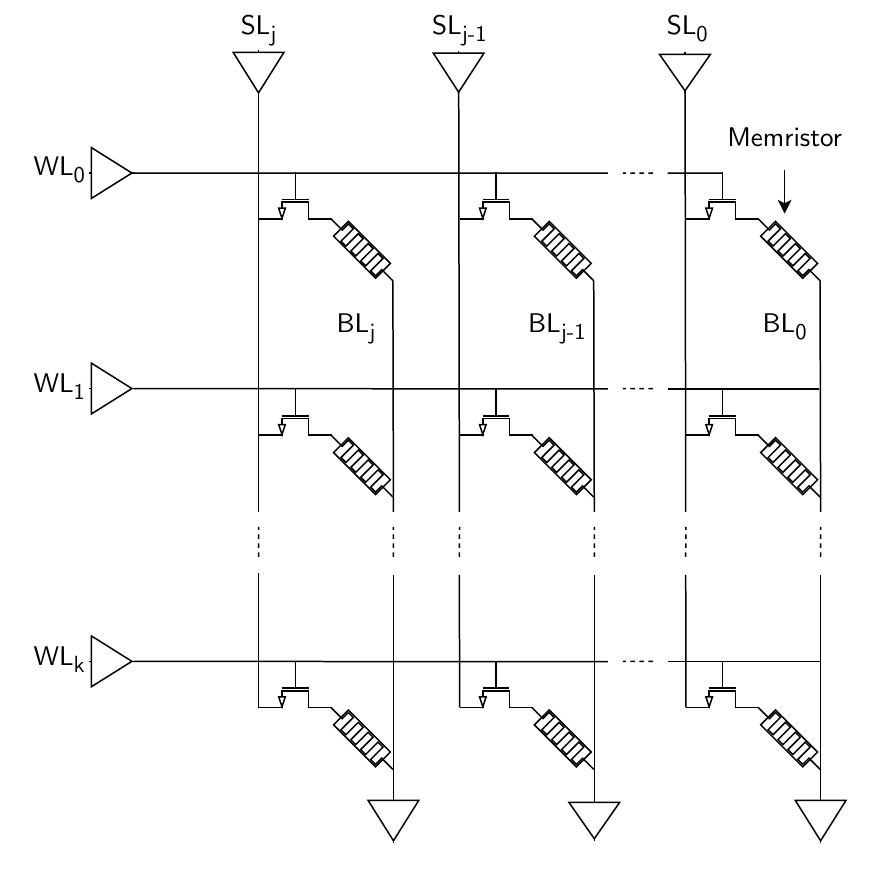}
    \caption{Basic structure of a 1T1R crossbar array.}
    \label{fig:1t1r-crossbar}
\end{figure}

The memristor is a two-terminal element (see Figure~\ref{fig:1t1r-crossbar}) that behaves like an ordinary resistor at a given instance~\cite{chua1971memristor}. It is \textit{non-volatile} and can either have a Low-Resistance State (LRS) or a High-Resistance State (HRS). These states depend on the voltage applied to one of the terminals and the duration over which this is done. It can be used as a non-volatile memory element, where HRS and LRS usually denote logic states `0' and `1', respectively. The processes of changing the resistance from LRS to HRS and HRS to LRS are called \textit{reset} and \textit{set}, respectively. As an emerging technology, memristor offers simplicity, fast switching speeds, ultra-low power consumption and high integration density~\cite{wang2020recent}. On top of that, the memristor is CMOS-compatible and shows the potential to overcome the von-Neumann bottleneck and sizing problem of transistors~\cite{li2021memristive}.

It is most common to configure memristors in a crossbar structure due to its simplicity.
A memristor crossbar is traditionally used as a memory structure to, for example, replace the traditional SRAM. This is also referred to as the Resistive-RAM (RRAM). Additionally, the structure may be used as an accelerator to drive neuromorphic applications by means of Vector-Matrix Multiplications (VMM)~\cite{zhang2019aging}. 

There are many different crossbar configurations, which are referred to as \textit{n-element-m-resistor} (nXmR) arrays; one crossbar bitcell consist of \textit{n} element(s) (such as, transistors or diodes) and \textit{m} memristor(s). 
The most basic and area-efficient configuration is 1R, which only consists of memristors. However, its difficulty lies in selecting individual bitcells: without a specified selector, current sneak paths are induced. This problem is avoided by implementing crossbars with an additional element per bitcell, such as by using 1T1R (1-transistor-1-memristor) (see Figure~\ref{fig:1t1r-crossbar}), which is the most popular and widely used structure.
It enables easy selection and programming of the bitcells for in-memory (in-situ) computations using Word Lines (WL) and Selection Lines (SL)~\cite{li2021memristive}. Next to that, the transistor and memristor elements can be stacked, allowing greater density. It is also possible to construct crossbars using more (memristive) elements. An example of this is the 2T2R crossbar proposed in~\cite{singh2022referencing}.

\subsection{Memristor-based hardware security}
\label{sec:memristive-security}

Besides their energy efficiency, memristors intrinsically show stochastic behavior~\cite{bai2014study} like resistance variability~\cite{emrl_res_material,dai2021audio}, probabilistic switching~\cite{sun2022physical,cai2022hyperlock} and Random Telegraph Noise (RTN)~\cite{huang2012contact}. On top of that, they also exhibit Device-2-Device/Cycle-2-Cycle (D2D/C2C) variations~\cite{yang2021cryptographic, cai2022hyperlock}, which increase the difficulty of performing side-channel analysis as they increase the noise levels in traces and make it harder to perform profiling attacks. With the aforementioned stochastic properties, Physical Unclonable Functions (PUFs), True Random Number Generators (TRNGs), and chaotic circuits can be built depending on the employed resistive material~\cite{pang2019memristors,lv2021application, sun2022physical, wang2019finite, liu2022dynamic}. These security schemes are used for authentication, key-generation and encryption. 

In terms of \textit{encryption}, a memristor-based block cipher can offer the following advantages: 
A memristor crossbar allows reconfigurability, which can be used to update the cipher substitution boxes (SBs) for masking at run-time to protect against side-channel attacks~\cite{gross2016domain}. In this case, the cipher output should be adjusted accordingly to maintain the correct functionality. The non-volatile nature of such crossbars allow the SBs to retain their value, which allows us to power off the design to save energy. 
On the other hand, a CMOS-only implementation will be using e.g., an SRAM or logic implementations of the SBs. SRAM-based solutions allow reconfiguration but have less variation and cannot be powered off, whereas logic implementations (such as and-or trees) cannot be modified. Another option could be to have embedded flash-based SBs, however, these memories are more power consuming than memristor-based ones (such as, RRAM) during active mode~\cite{oboril2015evaluation}.

\subsection{Related work}
\label{sec:related-work}

Sun et al.~\cite{sun2022physical} propose a memristor-based PUF that prevents attacks using triggered solubility when necessary. This is achieved by using water-assisted transfer printing. Cai et al.~\cite{cai2022hyperlock} use a randomly-initialized memristor crossbar to perform VMM for creating hypervectors as a means for encryption. It relies on crossbar non-idealities and C2C variations. However, for decryption, the authors propose a neural network. Similarly, in~\cite{yang2021cryptographic}, a 1T1R crossbar is used to store plaintext and perform in-situ XOR operations with key bits for encryption. The key bits are generated using the subthreshold slope of each transistor. These vary intrinsically and hence function as a PUF. Two other works discuss the concept of key-less encryption, using memristors as a source of entropy~\cite{wang2019finite, cambou2020cryptography}. 
Khedkar et al.~\cite{khedkar2014towards} propose a memristor-based AES design with the aim of protecting against differential power attacks. Each AES SB is implemented using a neural network having 8 inputs, 1 output, and a hidden layer of 48 neurons, which is not suitable for heavily resource-constrained edge devices. However, it should be noted that energy efficiency was not the aim of their work.

Despite these promising schemes, many of them are not necessarily lightweight, especially the above-mentioned block cipher.
In addition, the majority of these solutions require frequent operational switching of memristors, which should be avoided due to the increase in energy consumption and decrease in lifetime of the memristive device~\cite{zhang2019aging}. 
In brief, there is still a void in literature when it comes to a lightweight memristor-based block cipher that also supports energy-efficient side-channel protection.

\ifnum\journal>0
\fi

\section{Towards memristor-based encryption}
\label{sec:towards-mem-based-encryption}

\subsection{Looking for the right cipher}

In order to showcase the possibility of using memristors in lightweight block encryption, we refer to existing literature to find a suitable encryption scheme that can serve as a proof of concept.
Multiple extensive literature reviews on the state-of-the-art lightweight cryptography already exist~\cite{joseph2021analysis,thakor2021lightweight,naser2022systematic}.
We evaluate the available ciphers by looking at the \textbf{throughput}, \textbf{power/energy consumption}, \textbf{design simplicity}, and most importantly, their potential \textbf{compatibility with memristors}. These include: PRESENT, RECTANGLE, SIMON, SPECK, GIFT, SLIM, $\mu^2$, ANU-II, NLBIST, Piccolo and BORON.

In terms of throughput and energy, Speck, Simon and GIFT show the best results. Regarding simplicity, GIFT outperforms the other two~\cite{joseph2021analysis}. 
Moreover, its simple structure (mainly consisting of substitution boxes) enables a straightforward crossbar implementation. More specifically, a memristor crossbar can be composed in a way such that a GIFT encryption round can be performed by only a single \textit{read} action (as will be discussed in Section~\ref{sec:proposed-solution}). For these reasons, the GIFT cipher is used as a reference and inspiration towards implementing a lightweight memristor-based encryption block suitable for next-generation edge devices. The next section will briefly explain the working principle of GIFT.

\subsection{The GIFT cipher}
\label{sec:gift-cipher}

\begin{figure*}[!t]
    \includegraphics[trim={0cm 0.2cm 0cm 0.1cm},clip,scale=0.6]{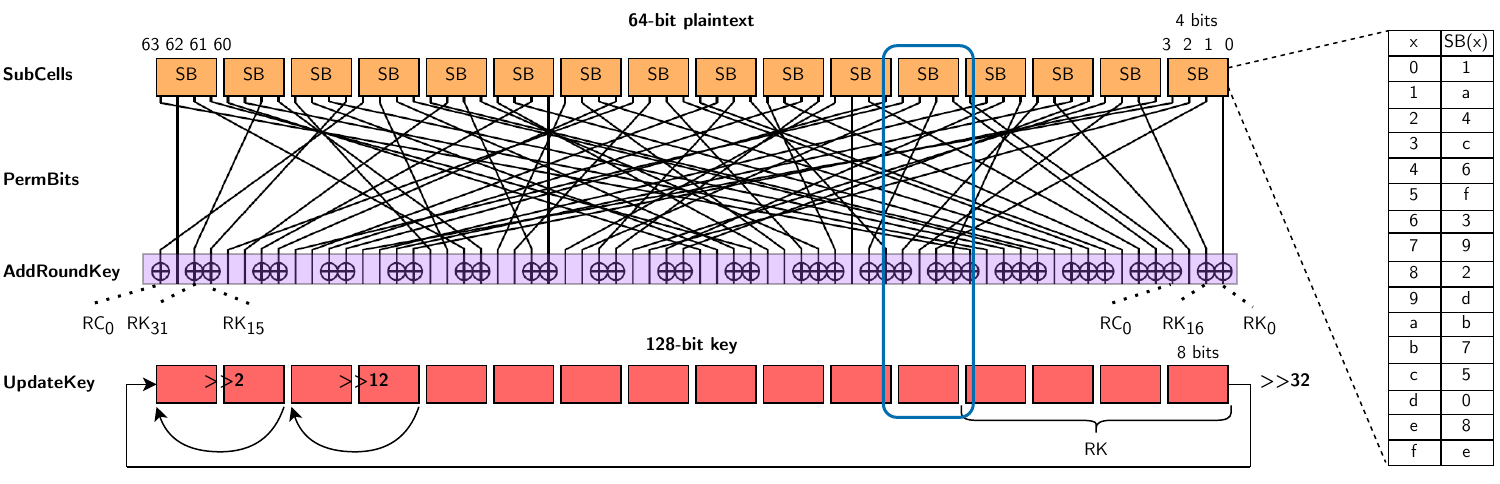}
    \centering
    \caption{Architecture of GIFT-64~\cite{banik2017gift}. The encircled section (in blue) denotes a slice of a single round of encryption. The substitution specification of a 4-bit SB is shown on the right.}
    \label{fig:gift-encryption}
\end{figure*}

Similar to the known standardized algorithms such as AES, SKINNY and PRESENT, GIFT is based on substitution-permutation network, in which the plaintext nibbles are replaced with other values, followed by rearrangement. The cipher is inspired by its predecessor PRESENT but has improved security and efficiency~\cite{banik2017gift}. 
The GIFT family consists of two members: GIFT-64 and GIFT-128. The former takes in 64 bits and uses 28 rounds while the latter encrypts 128 bits using 40 rounds. Both versions use a 128-bit encryption key. Figure \ref{fig:gift-encryption} shows the cipher architecture. A round of GIFT consists of three basic operations:

\begin{enumerate}
    \item \textbf{SubCells}: The plaintext nibble is substituted with a 4-bit sequence through an SB.
    Figure~\ref{fig:gift-encryption} shows the substitution specification for such a 4-bit SB. 
    \item \textbf{PermBits}: The output of the SB is permuted. Each of the 4 bits is re-routed through hard-wiring.
    \item \textbf{AddRoundKey}: For the 64(128)-bit version a 32(64)-bit Round Key (RK) is extracted from the \textit{key state}, which is partitioned into 2(4) 16-bit words. 
    Two bits of each of the permuted nibbles (i.e., two LSBs of the nibbles in the case of GIFT-64, and middle two bits in the case of GIFT-128) are XOR-ed with two bits from the RK.
    Figure~\ref{fig:gift-encryption} illustrates this for a single round of GIFT-64.
\end{enumerate}

Since only two key bits per nibble are used when performing the XOR operations, each RK is updated and shifted after every round to ensure that the other part of the key state is used. This is done by performing a 32-bit right rotation. This is followed by a 2-bit and 12-bit right rotation performed on the two MSB bytes and the two bytes thereafter, respectively.  
Figure~\ref{fig:gift-encryption} depicts the complete key state update. 
In addition to the RK, there also exists a 7-bit Round Constant (RC), which is applied to bit positions $n-1$, $3$, $7$, $11$, $15$, $19$ and $23$, where $n = 64, 128$. After every round, the RC is updated by means of a rotational left shift followed by two XOR operations between the new LSB, MSB and `1'. The RK and RC schedules are the same for both versions of the cipher.

\section{Proposed Solution}
\label{sec:proposed-solution}

\subsection{Design overview and assumptions}
\label{sec:assumptions}

\begin{figure}[!t]
    \includegraphics[trim={0cm 0cm 0cm 0cm},clip,scale=0.23]{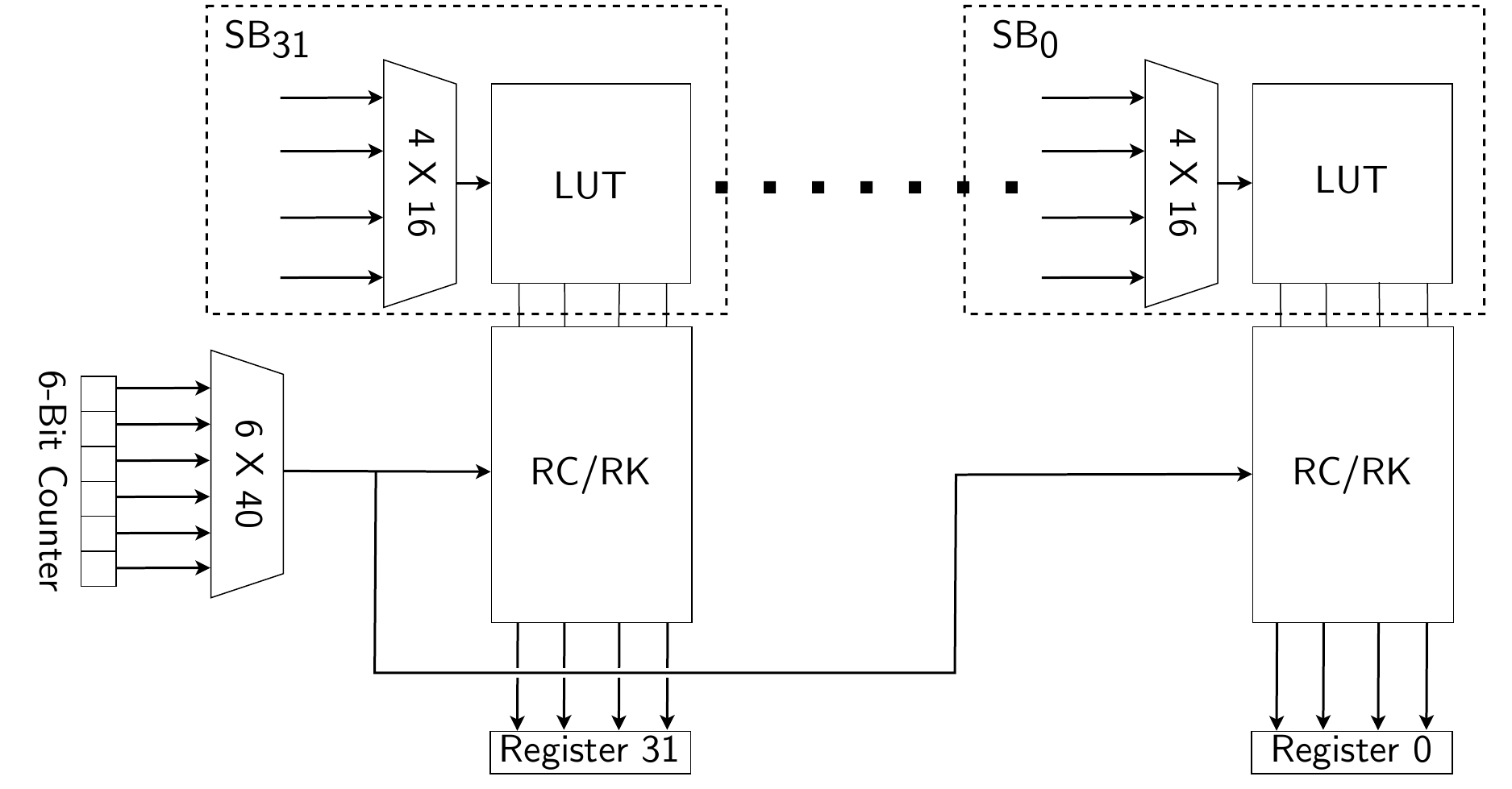}
    \centering
    \caption{Top view of the memristor-based GIFT cipher.}
    \label{fig:top-gift}
\end{figure}

The general idea behind our approach is to take advantage of the bit-slicing topology of GIFT and compress all the operations shown in Figure~\ref{fig:gift-encryption} into \textit{one} lightweight module made of a memristor-based crossbar.
As stated earlier, GIFT is a natural candidate for a crossbar-based implementation due to its simple structure.
For example, the first operation, i.e., \textit{SubCells}, is realized using 4-bit SBs, which are normally implemented using Look-Up Tables (LUTs). These LUTs can in turn be implemented using memristor crossbars.
A memristive crossbar structure is preferred over cascaded Memristor-based Logic Gates (MLGs) due to the possibility of high-density and in-memory computations. Moreover, a crossbar structure mimicking LUTs eliminates frequent writing/reset operations on memristors, which can introduce reliability issues and shorten lifetime of these components~\cite{zhang2019aging}.
Figure~\ref{fig:top-gift} shows the top view of the proposed design. To the best of our knowledge, no other work has proposed a memristor-based lightweight block cipher at the time of writing.

The key aspects of our approach are as follows:

\begin{itemize}
    \item Mapping the three GIFT operations and key scheduling to a memristor crossbar makes it possible to execute a GIFT encryption round in a single \textit{read} operation.
    \item Only at the start of an encryption session would a \textit{write} operation be required to program the constant/key-bit values.
    \item Each crossbar implementation covers the 40 encryption rounds, i.e., the same hardware is reused for each round.
    \item By minimizing switching activity and by mapping all the operations to the crossbar, lesser energy consumption and footprint is achieved.
\end{itemize}

Before diving deeper into the design, a couple of assumptions must be established. Firstly, in this work, the 128-bit GIFT version is considered. However, our approach also applies to the 64-bit version. Secondly, the encryption key needs to be created and exchanged beforehand between the transmitter and the receiver to perform the encryption. 
This prior key exchange is considered outside the scope of the paper and has been already addressed by prior work. For instance, in the case of a medical-implant edge application, a body-coupled-communication-based protocol can be used for this purpose~\cite{siddiqi2021securing}.

\subsection{Substitution box} 
\label{sec:s-box}

\begin{figure}[!t]
    \includegraphics[trim={0cm 0cm 0cm 0cm},clip,scale=0.4]{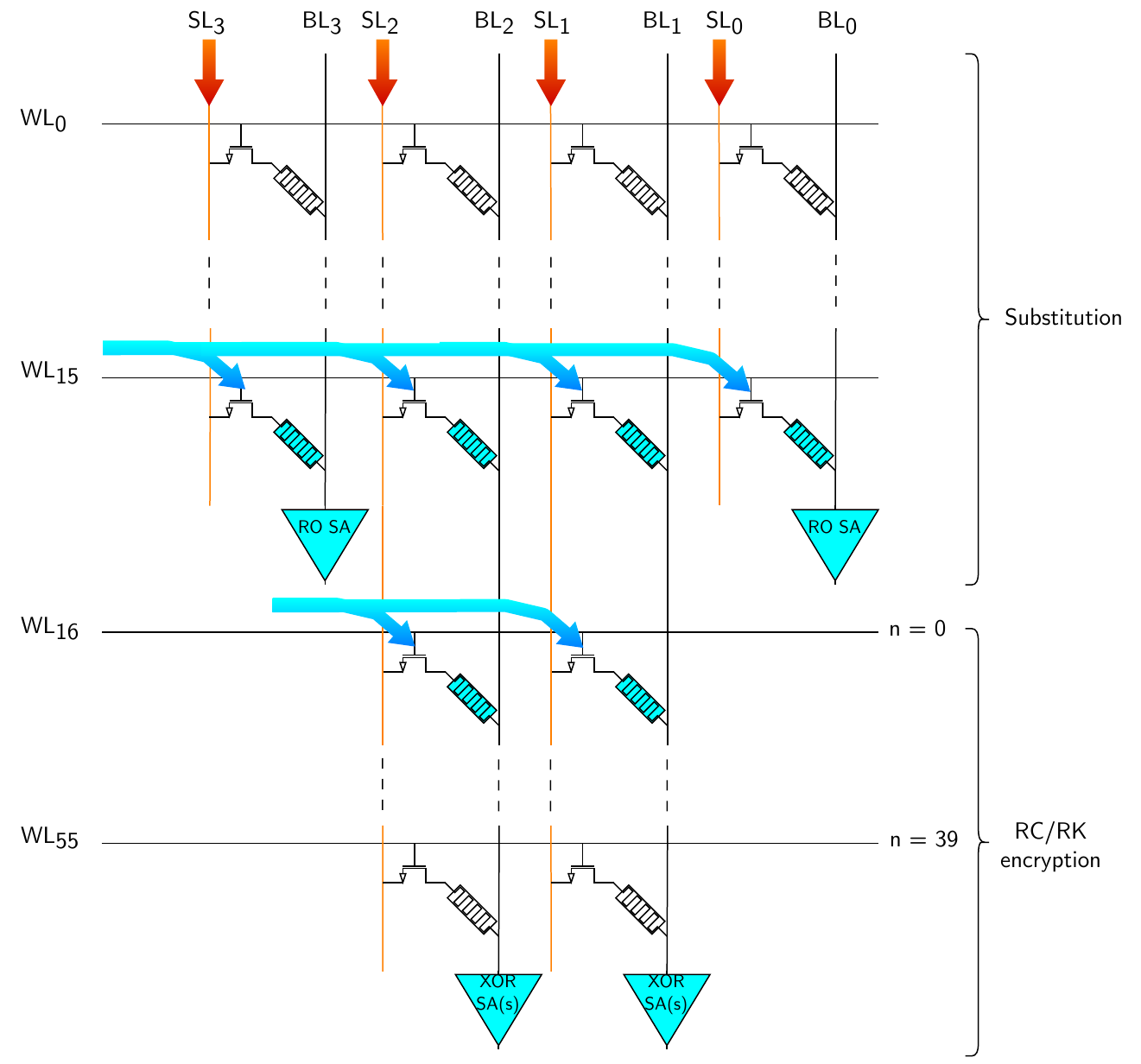}
    \centering
    \caption{The slice architecture of the 1T1R GIFT cipher and operation sequence corresponding to a round of encryption.}
    \label{fig:1t1r-gift}
\end{figure}

Regarding the memristor-based LUT implementation of the SB, there are multiple options for the crossbar structure: Passive crossbar arrays (1R) are an attractive solution for high-density and low-power integration. However, they have weaknesses such as current sneak paths and floating state issues during reading and writing operations. 
Furthermore, this leakage becomes more severe when the number of cells increases~\cite{taherinejad2021sixor}.
An established solution to this problem is the nTnR structure. 
The most commonly used structure is 1T1R, but 2T2R has also shown significant benefits~\cite{singh2022referencing}.
However, as the name suggests, 2T2R is twice the size of 1T1R. 
Mapping an SB LUT onto a 1T1R crossbar results in the structure shown in Figure~\ref{fig:1t1r-gift}. 
The SB unit can have 16 possible 4-bit input values, and the internal LUT has 16 rows of four memristor/transistor pairs, with each memristive cell representing 1 bit of the SB output. 
Given a 4-bit input, an address decoder selects one of the 16 rows (corresponding to Figure~\ref{fig:gift-encryption}) by applying a voltage to the word line (WL) connected to the transistor gates of that row, which enables the respective memristive cells. The details of the crossbar operation will be provided in Section~\ref{sec:round-enc}.

\subsection{XOR operation}
\label{sec:xor}

The second operation of the GIFT encryption is \textit{AddRoundKey}, which is performed using XOR. 
Finding the appropriate MLG for such an operation is crucial for achieving low energy and area overheads. 
As a result, the following notable MLG architectures from literature were considered: Memristor-Aided Logic (MAGIC)~\cite{kvatinsky2014magic}, Single-Cycle In-Memristor XOR (SIXOR)~\cite{taherinejad2021sixor}, Scouting Logic~\cite{xie2017scouting}, 1T1R In-Situ Boolean Logic~\cite{wang2016functionally}, Stateful 1T1R NANDs~\cite{shen2019stateful}, Material Implication Memristor Logic~\cite{kvatinsky2013memristor}, Memristor-Ratioed Logic (MRL)~\cite{kvatinsky2012mrl}, CMOS-Like Logic~\cite{vourkas2012novel}, Parallel Input-Processing Memristor Logic~\cite{papandroulidakis2014boolean}, and Dual Sense Amplifier Crossbar Logic (DSA)~\cite{jain2017computing}.

Many of these MLGs are limited due to (1) cascading problems, such as requiring signal restoration between stages, etc.; (2) destructiveness, i.e., the follow-up operation requires all memristors to be initialized again by means of a write operation; and (3) a long sequence of operations. MLGs implemented in a constrained environment should have very low power consumption and acceptable delay. MLGs that require many sequential resets may, therefore, be less desirable. As of now, the usage of standalone MLGs such as SIXOR, MAGIC, and MRL is too inefficient for encryption in next-generation edge devices~\cite{galvan2022memristive}. On the other hand, crossbar solutions show good potential for parallel in-memory computing applications. Moreover, they can be easily made compatible with the SB-unit crossbar. We thus employ both Scouting Logic and DSA to implement XOR in this work:

\subsubsection{Scouting-Logic-based XOR (\slv)}

\begin{figure}[!t]
    \includegraphics[trim={0cm 0cm 0cm 0cm},clip,scale=0.63]{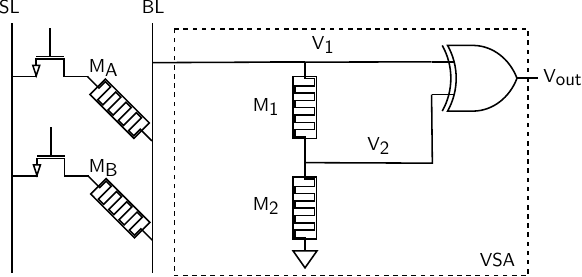}
    \centering
    \caption{\slv: VSA-based Scouting Logic XOR.}
    \label{fig:VSA}
\end{figure}

The Scouting Logic family~\cite{xie2017scouting} offers OR, AND, and XOR operations and is crossbar compatible. It is non-destructive as the resistive states are preserved across multiple operations. Hence, no switching between resistive states is necessary, which is power-efficient. Moreover, a single logic operation only requires one step, in which no initialization and restoration are needed as it primarily consists of a reading operation.
The core of the Scouting Logic design varies depending on the employed sensing scheme, i.e., voltage (VSA) and current sense amplifiers (CSA). 
It has been shown in~\cite{galvan2022memristive} that CSA has higher area and power consumption overheads than VSA. However, it is faster than VSA. Since performance is not a bottleneck for the targeted edge applications, VSA is employed in this work for the \slv design (see Figure~\ref{fig:VSA}).

VSA generates a reference using additional memristors M$_1$ and M$_2$. It also uses a CMOS XOR gate as a threshold function to determine the proper output pulse based on the generated reference. When a read pulse is applied from SL to M$_A$ and M$_B$ (which contain the two inputs for the XOR operation), a current determined by the equivalent input resistance will flow from the Bit Line (BL) to the VSA. Depending on whether the current (or resulting voltage) is higher or lower than the gate threshold, the output will be either `0' or `1'.

\subsubsection{DSA-based XOR (\dsa)}

In DSA~\cite{jain2017computing}, complex gates, such as bitwise XOR, are composed using two sense amplifiers (SAs).
The DSA scheme is meant for crossbar operations, and due to its simplicity, it is able to perform the operations using a single cycle only. However, this comes at the cost of using two SAs and a single 2-bit NOR gate.

\begin{figure}[!t]
    \includegraphics[trim={0cm 0cm 0cm 0cm},clip,scale=0.6]{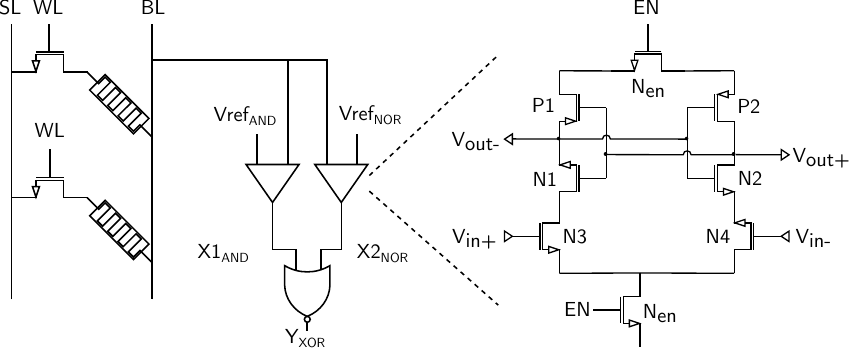}
    \centering
    \caption{\dsa: DSA-based XOR with the expanded voltage-based SA (right).}
    \label{fig:xor-crossbar}
\end{figure}

By combining the reference currents of the NOR and NAND SAs, an XOR operation is realized, following the equation $Y_{\text{XOR}} = X1_{\text{AND}} \; \textbf{NOR} \; X2_{\text{NOR}}$, as shown in Figure~\ref{fig:xor-crossbar}, where $X1$ and $X2$ are the outputs of the respective SAs.

In \cite{jain2017computing}, current mirrors are used in the crossbar to drive the SAs. For the design of the DXOR-GIFT, a voltage-based SA is used to eliminate the need for current mirrors, resulting in higher energy efficiency. The schematic for the voltage-based SA is adapted from the SA proposed in \cite{jain2017computing}, with minor alterations to target only the XOR functionality (see Figure \ref{fig:xor-crossbar}).

\subsection{Permutation}
\label{sec:permutation}

Considering the permutation scheme in~\cite{banik2017gift}, which is also depicted as the web of connections in Figure~\ref{fig:gift-encryption}, each $n^{th}$ bit of every nibble is connected to the $n^{th}$ bit of another nibble that drives the SB of the next round. For example, the $1^{st}$ bit (i.e., the second output) of SB$_0$ (the rightmost SB in the first round) is connected (permuted) to the second input of SB$_4$ during the next round, which is also encircled in Figure~\ref{fig:gift-encryption}. Thus, conceptually, SB$_0$ of round 1 drives SB$_4$ of round 2 with an XOR operation in between. However, there is no difference between the SBs of each round since they are all identical. This means that one can consider the $n^{th}$ output of SB$_n$ to simply be fed back to the $n^{th}$ input of the same SB.

Figure \ref{fig:gift-encryption} shows that it is always the same bits in each nibble that are involved in XOR operations. 
The only thing that changes is the value of the key bit after every \textit{UpdateKey}. Knowing this and the fact that every bit in the nibble is routed to the same relative position, it is possible to anticipate the RK and RC bit values that are used in the XOR operations of every round if the encryption key is known. More specifically, instead of performing RK/RC updates every round, these values are pre-computed (offline) and arranged at their relative bit positions within the nibble after following the permutation table. During the initialization of the encryption module, these values are uploaded only once to the memristors in WL$_{16}$ to WL$_{55}$, which saves significant overhead due to the removal of the active key-schedule operation in its entirety.

\subsection{Encryption round}
\label{sec:round-enc}

The means of incorporating the permutation and key-scheduling approach are illustrated in Figure~\ref{fig:1t1r-gift}. In addition to the SB crossbar proposed in Section~\ref{sec:s-box} for \textit{substitution}, a $40\times 2$ 1T1R crossbar (or $40\times 3$ for the few nibbles with additional RC) is connected in series for \textit{RC/RK encryption}. As mentioned in the previous section, this 40-row crossbar (WL$_{16}$ to WL$_{55}$) contains the RK (and RC) bit values arranged according to the permutation table and the key schedule for the complete encryption of a 128-bit plaintext. This means that \textit{active} key scheduling is not required anymore.
For GIFT-128, only the middle two bits of each nibble are encrypted with the appropriate RK bits~\cite{banik2017gift}, as opposed to the first two bits of each nibble for GIFT-64. 
The XOR SAs (i.e., DSA-based or scouting-logic-based) are connected to the bottom of the crossbar. 
Figure~\ref{fig:1t1r-gift} also illustrates the flow of all the above operations for a single slice. An encryption round is as follows:

\begin{enumerate}
    \item Based on the SB input and the mapping presented in Figure~\ref{fig:gift-encryption}, the corresponding WL is driven by a voltage pulse, upon which the NMOS switches in that row are closed, and the respective memristors are selected. In addition, the RC/RK memristors corresponding to the first round are selected. The activated rows are shown using blue arrows in Figure~\ref{fig:1t1r-gift}.
    \item A read pulse is generated for each SL (red arrows in Figure~\ref{fig:1t1r-gift}), allowing current to flow through the memristors and the BL, into the XOR and read-out (RO) SAs.
    \item With the desired rows selected, an XOR operation is performed between the top and bottom parts for bitcells 1 and 2, respectively. Depending on the resistive states of these bits, a `0' or `1' pulse is generated at the output. Bitcells 0 and 3 are read out without performing XOR.
    \item All the above steps are performed within the same cycle, and the output is temporarily stored in an output register.
    \item In the next round, the register output is fed back again to the input of the same SB.
\end{enumerate}

\subsection{Crossbar address decoders}
\label{sec:address-decoders}

The purpose of the address decoders in Figure~\ref{fig:top-gift} is to drive the desired WLs in the crossbar, resulting in the selection of the memristors embedded in the corresponding row. 
These decoders can add significantly to the design overhead since our GIFT-128 design requires 32 4-to-16 address decoders and a single 6-to-40 address decoder (see Section~\ref{sec:rc-rk-selector}). 
As a result, different designs were considered for the optimized implementation of these decoders.
For example, one approach is to decode the bits by means of a sequence of PMOS and NMOS transistors connected in series, where each one of them corresponds to the expected input bit. 
The disadvantage of such a circuit is that it requires precise sizing of the transistors. Moreover, the voltage drop caused across a transistor becomes an issue when considering the transistor overdrive. Consequently, a repeater is required to provide enough input drive for the WLs.
A better solution is to use a complete logic scheme that allows for the use of the smallest possible transistor sizes. 
A second approach is based on NAND/NOR trees, such as the one proposed for decoding addresses in SRAMs~\cite{mishra2014novel, singh2017low}. This approach is relatively lightweight, reliable, and can also be applied to memristor crossbar arrays, which is why it is used in this work.

\subsection{RC/RK selector}
\label{sec:rc-rk-selector}

The entire architecture requires a single RC/RK selector for selecting the WLs in all the 32 RC/RK units shown in Figure~\ref{fig:top-gift}. 
The goal of this selector is to select a WL for every round, from round 1 to 40.
One approach is to use a shift register for selecting each WL after every bit shift. However, a 40-bit shift register is costlier than using a 6-bit counter that drives a 6-to-40 address decoder. 
As a result, the latter option is used for our purpose.

\section{Results}
\label{sec:results}

\subsection{Experimental setup}

Both the 1T1R-GIFT designs, i.e., the ones based on the DSA-based XOR (\dsagift) and scouting-logic-based XOR (\slvgift), were implemented as SPICE netlists using Cadence Virtuoso. Cadence Spectre was used for design simulation/verification and power measurements. The designs were realized using the TSMC 40 nm library.
In the original work \cite{banik2017gift}, the GIFT implementation runs at a frequency of 10 MHz. The same is done for the 1T1R implementations for consistency.
Table \ref{tab:designParam} summarizes the design parameters for these implementations.

\begin{table}[!t]
\centering
\caption{Design parameters}
\label{tab:designParam}
\begin{threeparttable}
\begin{tabular}{@{}ll@{}}
\toprule
Technology node        & 40 nm      \\
Technology   & RRAM (HfO$_2$$^*$) \\
Operating voltage & 0.9       \\
Operating frequency   & 10 MHz    \\
\bottomrule
\end{tabular}
\begin{footnotesize}
\begin{tablenotes}
 \item[*] JART VCM v1b model from~\cite{emrl_res_material}.
  \end{tablenotes}
  \end{footnotesize}
 \end{threeparttable}
\end{table}

\subsection{Implementation}

The 1T1R-GIFT SPICE netlist entails a single GIFT-128 slice, as highlighted in Section~\ref{sec:assumptions}. The 1T1R model employed in our implementation is adapted from \cite{singh2022referencing}, which uses the HfO$_2$-based memristor from \cite{emrl_res_material} and considers non-idealities such as wire resistances and capacitances.
All the CMOS-based logic used in the designs is implemented using the cells provided by the TSMC 40 nm standard library. In general, minimal size is used for the gates. However, the gates in the final stages of the decoders are four times larger to ensure sufficient drive strength for the WLs.

\subsubsection{\slvgift} 
In \cite{xie2017scouting}, while operating Scouting Logic VSA, the resistive states of the memristors are retained. Hence, as with all other memristors in this design, they only need to be programmed once. For the voltage divider to work properly, the acting memristors need to be scaled accordingly to guarantee an output of (at least) 0.6 V$_{DD}$ and 0.4 V$_{DD}$ for logic `1' and `0', respectively. Since this work utilizes a different structure and technology, the scaling rules stated in \cite{xie2017scouting} cannot be applied. For example, the original Scouting Logic VSA requires M$_1$ and M$_2$ to be $2\times LRS$ and $2.5\times LRS$, respectively. However, when doing so for the proposed cipher, the small resistive values result in V$_1$ and V$_2$ not crossing 0.20-0.30 V, which is way below the required threshold value of 0.45 V. Moreover, keeping the 1:1.25 ratio between the two memristors in Scouting Logic VSA results in too large of a margin between V$_1$ and V$_2$, causing V$_2$ to stay below the threshold and resulting in operational failure.
Using the simplified model of the employed memristor crossbar, the proper resistor values were determined: it was found that \slvgift performs well when using 2 k$\Omega$ and 250 k$\Omega$ for M1 and M2, respectively.
Since the XOR operations are only performed on the middle two bits of each nibble, the remaining LSBs and MSBs just need to be read out using a read-out (RO) SA. For this purpose, Scouting Logic VSA is downsized to just a single memristor, M$_1$, which is programmed to 550 k$\Omega$, and the XOR gate in the SA is replaced by an OR gate.

\subsubsection{\dsagift}

For the AND and NOR SAs in the \dsa sensing schemes, constant voltage references of 0.45 V and 0.43 V are used, respectively. Regarding the read-out of the LSBs and MSBs, only a single read-out SA suffices, for which a reference of 0.43 V is used.

\subsection{Energy- and area-efficiency analysis}

\begin{figure}
	\centering
	\subfloat[\slvgift]{\label{fig:slv-overheads}{\includegraphics[trim={1.5cm 1.2cm 0cm 0.7cm},clip,scale=0.6]{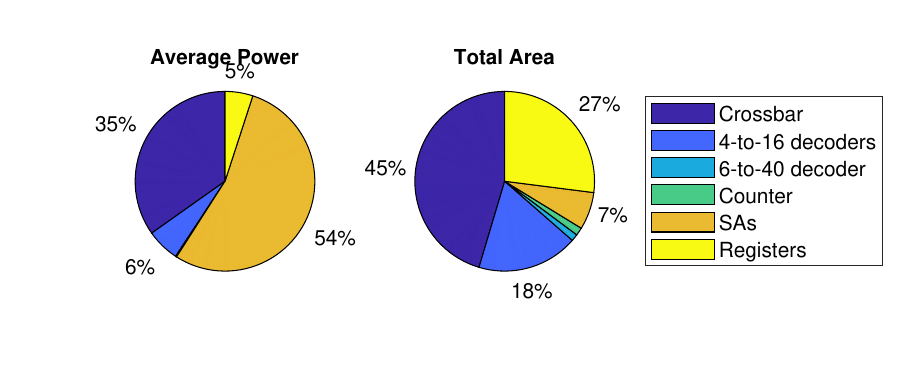} }}%
	\qquad
	\subfloat[\dsagift]{\label{fig:dsa-overheads}{\includegraphics[trim={1.5cm 1.2cm 0cm 0.7cm},clip,scale=0.6]{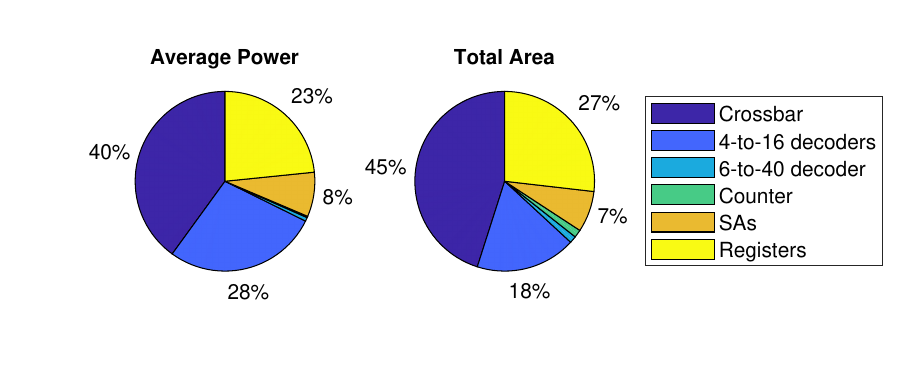} }}%
	\caption{Average-power and total-area breakdown}%
	\label{fig:overheads}%
\end{figure}

The implementation results are summarized in Table \ref{tab:Implementation_opt_res}. 
It can be seen that both \slvgift and \dsagift consume almost the same amount of area. However, \slvgift has more than $4 \times$ the power consumption of \dsagift.
The breakdown of the average power and total area is illustrated in Figure~\ref{fig:overheads}. It can be seen that the scouting logic SAs are the bottleneck in the case of \slvgift, whereas power consumption is relatively evenly spread in the case of \dsagift. Overall, \dsagift outperforms \slvgift and hence can be considered a better approach for implementing lightweight block ciphers.

\begin{table}[!t]
\centering
\caption{Implementation Results}
\label{tab:Implementation_opt_res}
\begin{adjustbox}{width=.95\columnwidth}
\begin{threeparttable}
\begin{tabular}{@{}llll@{}}
\toprule
                       & \slvgift & \dsagift & CMOS-GIFT$^{\S}$~\cite{banik2017gift}  \\ \midrule
Average Power ($\mu$W) & 257.6$^*$          & 60.38$^*$ & 116.6 \\
Energy (pJ)            & 1030.4             & 241.52 & 478.1 \\
Area (mm$^2$)          & 0.0034             & 0.0034  & --    \\
%Delay per round (ns)   & \hl{TODO: JA}      & \hl{TODO: JA} &  1.85 \\
Latency$^{**}$ (us)    & 4 & 4 & 4       \\ 
\bottomrule
\end{tabular}
 \begin{footnotesize}
 \begin{tablenotes}
 \item[] $^*$ Extrapolated for 32 slices $\vert$ $^{**}$ Duration of 40 encryption rounds (10 MHz clock) $\vert$ $^{\S}$ Using STM 90 nm library $\vert$ `--': Lacking information.
 \end{tablenotes}
 \end{footnotesize}
 \end{threeparttable}
\end{adjustbox}
\end{table}

\subsection{Discussion}

Our analysis indicates that memristors show significant promise in terms of their use as building blocks for lightweight block ciphers. 
In terms of energy, processing, and area overheads, the results of the 1T1R implementation are roughly twice as good as those of the prior 90-nm CMOS-only implementation~\cite{banik2017gift}, as shown in Table~\ref{tab:Implementation_opt_res}.
More importantly, this new approach to designing lightweight block ciphers offers additional low-cost protection against side-channel and template attacks, as highlighted in Section~\ref{sec:memristive-security}. 
Lastly, our analysis shows that \dsa is a preferable approach compared to the state-of-the-art MLGs available in the literature.
However, there may be room for improvement in the 1T1R-GIFT implementation depending on future advances in technology:

Researchers have pointed out the difficulty of creating long MLG chains due to the loss in drive strength over several gate stages and the destructive nature of the operations. Despite numerous solutions, it is still sub-optimal due to limitations related to the number of operands, the required number of repeaters, and the number of operations. Solving these problems would allow SB construction using a simple memristor-based AND-OR tree, thereby significantly reducing the footprint.

Another alternative could be to shift towards a 2T2R approach, as in \cite{singh2022referencing}. It would result in a twofold increase in area, but in-situ operations can be done using differential sensing, which doubles the sensing margin compared to \cite{xie2017scouting} and hence increases reliability. Also, the energy consumption with 2T2R operands is significantly lower.
Unfortunately, this structure does not support XOR operations yet. 
Besides, researchers are also exploring the possibility of \textit{stacking} memristor crossbars, thereby significantly reducing the architecture footprint. These are also referred to as \textit{3D vertical RRAMs}. Some successful proposals exist that exploit different memristor-transistor compositions to produce 1T1R pillars \cite{wu2020monolithic,ezzadeen2020ultrahigh}.

Lastly, the transistor of the 1T1R structure is a limiting factor in scaling down the crossbar. However, advancements in resistive switching materials and selection methods can help in solving this limitation.

So, can ultra-resource-constrained edge devices benefit from a memristor-based security solution? At the moment, it does seem so. On top of that, further advancements in RRAM technology can potentially unlock the full capability of memristor architectures and, consequently, make them a fitting building block for ultra-lightweight security applications.

\section{Conclusions}
\label{sec:conclusions}

In this paper, we proposed a lightweight implementation of the GIFT cipher using an RRAM-based architecture in a 1T1R configuration. The design was implemented using a 40 nm process technology. Not only do our scheme's area and power overheads compare favorably to those of a CMOS-only implementation, but it also allows for housing the substitution boxes (SBs) in an energy-friendly and non-volatile fashion. The reconfigurability of these SBs allows the cipher operation to be masked, providing protection against side-channel attacks at no significant additional cost.
Such a lightweight design can significantly contribute to securing small-form-factor edge devices for next-generation personalized healthcare.

% use section* for acknowledgement
\section*{Acknowledgment}

This work has been supported by the EU-funded projects SEPTON, SECURED, and CONVOLVE, with the grant agreement numbers 101094901, 101095717, and 101070374, respectively.

\bibliographystyle{IEEEtran}
\bibliography{References}

\end{document}